\documentclass[twocolumn,showpacs,showkeys,superscriptaddress]{revtex4}

\usepackage{amsmath}
\usepackage{amsfonts}
\usepackage{amssymb}
\usepackage{pbsi}
\usepackage[T1]{fontenc}
\usepackage{hyperref}
\usepackage{xcolor}
\usepackage{color}

\def\openone{\leavevmode\hbox{\small1\kern-3.8pt\normalsize1}}
\def\N{\leavevmode\hbox{ Z \kern-8 pt\normalsize{Z}}}
\def\openone{\leavevmode\hbox{\small1\kern-3.8pt\normalsize1}}
\def\openJ{\leavevmode\hbox{J \kern-9.5pt\normalsize J}}
\def\openS{\leavevmode\hbox{ S \kern-9.3pt\normalsize S}}

\newcommand{\bb}{\begin{equation}}
\newcommand{\ee}{\end{equation}}
\newcommand{\eqb}{\begin{eqnarray}}
\newcommand{\eqf}{\end{eqnarray}}

\begin{document}

\title{Bohm potential is real and its effects are measurable}

\author{Sergio A. Hojman}
\email{sergio.hojman@uai.cl}
\affiliation{Departamento de Ciencias, Facultad de Artes Liberales, Universidad Adolfo Ibáñez, Santiago 7491169, Chile.}
\affiliation{Departamento de Física, Facultad de Ciencias, Universidad de Chile, Santiago 7800003, Chile.}
\affiliation{Centro de Recursos Educativos Avanzados, CREA, Santiago 7500018, Chile.}
\author{Felipe A. Asenjo}
\email{felipe.asenjo@uai.cl}
\affiliation{Facultad de Ingeniería y Ciencias, Universidad Adolfo Ibáñez, Santiago 7491169, Chile.}
\author{H\'ector M. Moya-Cessa}
\email{hmmc@inaoep.mx}
\affiliation{Instituto Nacional de Astrofísica Óptica y Electrónica\\Calle Luis Enrique Erro No. 1, Santa María Tonantzintla, Pue., 72840, Mexico.}
\author{Francisco Soto--Eguibar}
\email{feguibar@inaoep.mx}
\affiliation{Instituto Nacional de Astrofísica Óptica y Electrónica\\Calle Luis Enrique Erro No. 1, Santa María Tonantzintla, Pue., 72840, Mexico.}

\date{\today}

\begin{abstract}
We analyze Bohm’s potential effects both in the realms of Quantum Mechanics and Optics, as well as in the study of other physical phenomena described in terms of classical and quantum wave equations. We approach this subject by using theoretical arguments as well as experimental evidence. We find that the effects produced by Bohm’s potential are both theoretically responsible for the early success of Quantum Mechanics correctly describing atomic and nuclear phenomena and, more recently, by confirming surprising accelerating behavior of free waves and particles experimentally, for instance.

\end{abstract}


\maketitle
 
 \section{Introduction}
 
A recently published article \cite{umul} deals with the reality of Bohm’s potential. In the last section of the paper, the sentence
{\it{``As a result, our analysis put forth that the term, named as the quantum potential, must be equal to zero.}}'' is, at least, misleading, if not outright wrong.

We believe it is important to remark that this conclusion is erroneous, as we demonstrate in the following sections of our manuscript, by considering theoretical as well as experimental results. In order to do this, we discuss some of the numerous findings published in articles related to the study of the Bohm potential and its effects, which may be traced back to almost a century ago in the seminal works of Madelung and of Bohm published in 1927 and 1952, respectively \cite{madel,bohm}.

Numerous authors published theoretical as well as experimental articles dealing with Bohm’s potential, touching upon different subjects including quantum mechanics, optics, field theory and general relativity
\cite{slep,skro,pleb,dewitt,vz1,mas1,har,mas2,mas3,sahthesis,hor,sah1,berry,sivichis2,sivichis,bloch,riv1,ah1,ah2,rivka,asenjoHojmanclasquandsper,sahfaz20201,sahfaz20202,impens,uaiinaoe, KGHojmanAsenjo,phol,holland,wyatt}. Most of the authors did not mention (or realized that) the fact that it was the non–-vanishing character of Bohm’s potential which produced the new (and sometimes counter--intuitive or surprising) phenomena.

A crucial issue about the Bohm potential is that some of the wave equation solutions have properties that depart from their classical (or point--like) counterparts. The new effects appear because waves are extended objects in contradistinction to the strictly local character of particles. The difference between classical (point--like) and quantum (wave--like) behavior is illustrated by confronting the WKB (approximated) treatment of a quantum problem (as compared) to the full quantum behavior of Schr\"odinger equation exact solutions or, equivalently, to the difference between ray (eikonal) approximation and wave optics treatments, for instance.

One simple way in which the wave-like behavior produces a striking effect that is impossible to get in the point-like (classical) limit, was found by Berry and Balazs in 1979 \cite{berry}. They showed that an accelerating Airy beam solves exactly the full quantum Schr\"odinger equation for a free (vanishing external potential) particle. This surprising result was later {\it{experimentally}} confirmed  using light beams \cite{sivichis} in 2007 and its acceleration control \cite{Cerda} in 2011, and electron beams \cite{bloch} in 2013. These phenomena could not have taken place for vanishing Bohm potential, as we show below. Thus, the Bohm potential for Airy beams is non-vanishing.

Similar to that solution, several others share this feature, producing a series of new and different phenomena. Below we show the theoretical foundations for the existence of a non--vanishing Bohm potential, to later discuss the physical effects of its presence, and how it produces new solutions for different wave equations.

\section{The Madelung--Bohm formulation of Quantum Mechanics}

Let us consider the usual way to represent the one--dimensional Schr\"odinger equations for a particle moving in a real external potential $V({x},t)$ in terms of complex wavefunctions $\psi=\psi({x},t)$ and $\psi^*=\psi^*({x},t)$,
\begin{equation}
-\frac{{\hbar}^2}{2m}\psi'' + V \psi - i \hbar \dot\psi = 0\, ,\label{schr1}
\end{equation}
and 
\begin{equation}
-\frac{{\hbar}^2}{2m} \psi^{* ''} + V \psi^* + i \hbar \dot\psi^* = 0\, ,\label{schr2}
\end{equation}
where $' \equiv \partial_x$ and $\dot\  \equiv \partial_t$.
The Madelung--Bohm \cite{madel,bohm} version of quantum mechanics is equivalent to Schr\"odinger's. It is written in terms of the polar decomposition of ${{\psi}} = A\exp\left(i S/\hbar\right)$, where the amplitude $A({ x},t)$ and the phase $S({ x},t)$ are real functions.
The two real Schr\"odinger equations, written in terms of these two new functions, are \cite{madel,bohm}
\begin{eqnarray}
\frac{1}{2m} \left(S'\right)^2 +V_B + V +\dot S&=&0\, , \label{HJB1} \\  \frac{1}{m}  \left(A^2 \, S'\right)'  + \left ({A^2}
 \right){\dot \ }&=&0\, . \label{cont}    
\end{eqnarray}
where the Bohm potential $V_B(x,t)$ is defined by
\begin{equation}
V_B \equiv -\frac{\hbar^2}{2m}\frac{ A'' }{A}\ . \label{VB}    
\end{equation}
The first equation \eqref{HJB1} is sometimes called the Quantum Hamilton--Jacobi (QHJ) equation for the (external) potential $V$. The quantum modification consists in the addition of the Bohm potential to the classical Hamilton-Jacobi equation. Besides, the second equation \eqref{cont} is the continuity (probability conservation) equation. Eqs.~\eqref{HJB1} and \eqref{cont} can be straightforwardly written for a three-dimensional space.

Interestingly, in one-dimension, the continuity equation \eqref{cont} is identically solved by defining the arbitrary wavefunction potential function ${f}={ f}({ x},t)$, such that 
\begin{equation}
 f'=A^2\, ,  \qquad
\dot{ f}=-\frac{A^2}{m} S'\ .   
\label{functionff}
\end{equation}

\section{Theoretical arguments and Experimental results}
In this section we analyze different theoretical topics and experimental findings that establish that Bohm potential may be different from zero. 

First of all, note that Planck's constant $\hbar$ appears {\it {only}} in Bohm potential in the Madelung--Bohm formulation of Quantum Mechanics, which is, of course, equivalent to the usual formulation. {\it{If Bohm potential vanishes identically, $\hbar$ would disappear in the formulation of Quantum Mechanics.}}
Even though this observation could debunk, by itself, the notion that in Quantum Mechanics Bohm’s potential can vanish identically, we will proceed to deepen our analysis further. 

There are many results, found independently,  that point out to the fact that classical and quantum dispersion relations are not equivalent \cite{slep,skro,pleb,dewitt,vz1,mas1,har,mas2,mas3,sahthesis,hor,sah1,berry,sivichis2,sivichis,bloch,riv1,ah1,ah2,rivka,asenjoHojmanclasquandsper,sahfaz20201,sahfaz20202,impens,uaiinaoe,KGHojmanAsenjo,phol,holland,wyatt}. 
It was recently found out \cite{asenjoHojmanclasquandsper} that all of these results have a common feature: the classical and quantum dispersion relations do not coincide because, in all of the cases considered (and in many other examples), {\it{the Bohm potentials do not vanish}}.

Furthermore, for a given external potential $V(x,t)$ the expressions for the Bohm potential $V_B(x,t)$ depend on the the wavefunction amplitude $A(x,t)= \sqrt{(\psi(x,t) \psi^*(x,t))}$. Therefore, one quantum problem, say the free particle, which is solved by infinitely many different wavefunctions, does not have a unique Bohm potential.
As a matter of fact, in a recent article \cite{sahfaz20201}, it is proved that not all external potentials $V(x,t)$ are compatible with vanishing Bohm potentials. No wavefunctions solutions to the Schr\"odinger equations for the Morse potential (or the P\"oschl--Teller potential, for instance) produce vanishing Bohm potentials.

The necessary and sufficient condition for an external potential to accommodate some (but not all) wavefunction solutions which have vanishing Bohm potential can be given in terms of the wavefunction potential function $f(x,t)$, defined through relations \eqref{functionff}. In a one-dimensional case,  for any system with vanishing Bohm potential,
the potential  $f$ function is given by
\begin{equation}\label{VB=0}
f(x,t)=\frac{a(t)^2}{3} x^3+a(t)b(t) x^2+b(t)^2 x+c(t),    
\end{equation}
for arbitrary functions $a(t)$, $b(t)$ and $c(t)$. In this case, the external potential $V=V(x,t)$ is given by
\begin{eqnarray}\label{VExt}
V=\ -\ \left( \frac{1}{2} \frac{m {\dot f}^2}{f'^2} + m \int_{\tilde x=0}^{\tilde x=x} \left(\frac{\dot f \dot f'}{f'^2}-\frac{\ddot f}{f'} \right) d\tilde x +\dot\mu \right), 
\end{eqnarray}
where $\mu(t)$ is an arbitrary function of time only, which produces a force $F=F(x,t)=\ -\ V'(x,t)$,
\begin{eqnarray}\label{F}
F= \left( \frac{1}{2} \frac{m {\dot f}^2}{f'^2} \right)'+ m \left(\frac{\dot f \dot f'}{f'^2}-\frac{\ddot f}{f'} \right)\ .
\end{eqnarray}
The free particle, the attractive and repulsive harmonic oscillators belong to this family, for instance. It is important to stress that this means that some (but not all) of the solutions to the aforementioned problems produce vanishing Bohm potentials.  Besides, any wavefunction potential function $f$ that does not fulfill \eqref{VB=0} can produce a non-vanishing Bohm potential for some external potential.
Details are given in Ref.~\cite{sahfaz20201}.

In  a different scenario, there are other well-known quantum solutions with non-vanishing Bohm potential.
The usual solutions to the $n$-bound state of the harmonic oscillator with mass $m$ and angular frequency $\omega$ may be written  for a wavefunction with an amplitude
\begin{equation}\label{An}
A_n(x,t)=\frac{1}{\sqrt{2^n\ n!}}\cdot \left(\frac{m \omega}{\pi \hbar} \right)^{\frac{1}{4}} e^{-\frac{m \omega x^2}{2 \hbar}}  H_n \left(\sqrt{\frac{m \omega}{\hbar}} x \right) \ ,
\end{equation}
and  phase $S_n(t)= - \left(n +{1}/{2}\right)\hbar\omega t$,  where $H_n$ are the Hermite polynomials. The Bohm potentials associated to these solutions are clearly different from zero and depend on $n$.
These solutions (with non--vanishing Bohm potential) are crucial in the construction of the well known and successful  ``Shell Model'' in nuclear physics. One cannot therefore, claim that solutions with non--vanishing Bohm potential are in some sense ``unphysical''. Something similar happens with the well known solution of the quantum Coulomb problem which accurately predicts the atomic spectra and transition probabilities and gives rise to understanding spectroscopy and the Periodic Table of Elements.

 On the other hand, a free solution may also produce a non-vanishing Bohm potential. The well-known  plane wave free particle solution $\psi_1(x,t)= A \exp(ik x -i\omega t)$, produces a vanishing Bohm potential, but there are other not so well known Airy function free particle solutions which produce non--vanishing Bohm potentials.
Let us consider the Berry and Balazs solution \cite{berry} given by
\begin{subequations}
\begin{align}
A(x,t)=& \mathrm{Ai} \left(\frac{\beta}{\hbar^{2/3}} \left(x-\frac{\beta^3}{4m^2} t^2 \right)\right),
\\
S(x,t)&=\frac{\beta^3 t}{2m}\left(x - \frac{\beta^3}{6m^2} t^2\right)\, ,
\end{align}
\end{subequations}
with a non--zero constant $\beta$. This solution of Schr\"odinger equation
for free particles, with $V=0$, has a non--vanishing Bohm potential which depends both in space $x$ and time $t$ given by
\begin{equation}
    V_B(x,t)=-\frac{\beta^3}{2m}\left(x-\frac{\beta^3}{4m^2}t^2\right)\, .
\end{equation}
This Bohm potential produces the constant acceleration $a_{\textrm{Airy}}$ experienced by the Airy wave packet 
\begin{equation}
   a_{\textrm{Airy}} = -\frac{V_B'}{m}=\frac{\beta^3}{2m^2}\, .
    \label{Bohmairy}
\end{equation}
This result gives rise to the velocity of the Airy package $v_{\textrm{Airy}} = p_{\textrm{Airy}}/m$ that may be computed from the momentum $p_{\textrm{Airy}}$ which is given by $p_{\textrm{Airy}}=\partial S(x,t)/ \partial x$.
Therefore, there is a solution to the free Schr\"odinger equation which has a constant acceleration given by \eqref{Bohmairy} in spite of being in the presence of a vanishing (external) force.
Of course, the most amazing feature of the Berry--Balazs result is that the acceleration of free beams has been experimentally detected in 2007 using light beams \cite{sivichis} and in 2013 using electron beams \cite{bloch}.

Finally, it has been shown in Ref.~\cite{sahfaz20202} that a non-vanishing Bohm potentials may  cancel some external potentials, allowing to have quantum solutions for particles  that behave as free classical particles even though  they are interacting with an external potential.

\section{Discussion}
We have clearly established by using both theoretical and experimental arguments, and examples, that the Bohm potential and its effects are real and measurable (even tough unknown or misunderstood almost a century after its definition).

The above discussed examples (and many others in the references) show that a non--vanishing Bohm potential has different kinds of effects in quantum mechanics, optics and wave propagation, in general. It is our belief that the recognition of the role that it plays in wave dynamics can bring up new and interesting insights in these different fields.


\end{document}